\documentclass[twocolumn,doublespacing]{bmcart}

\RequirePackage{natbib}
\RequirePackage{hyperref}
\usepackage[utf8]{inputenc} 
\usepackage{xr}
\usepackage[normalem]{ulem}
\usepackage{subfigure}
\usepackage{graphicx}
\usepackage{comment}
\usepackage{amsmath}
\usepackage{amsfonts}
\usepackage{amssymb}
\usepackage{algorithm}
\usepackage{algpseudocode}
\usepackage{booktabs}
\usepackage{tikz}
\usetikzlibrary[automata, positioning]
\usepackage{url}
\usepackage[flushleft]{threeparttable}
\usepackage{multirow}
\usepackage{caption}
\usepackage{xspace}
\usepackage{verbatim,soul}
\usepackage{adjustbox}
\usepackage{nccmath}
\usepackage{mathtools}
\usepackage{rotating}
\usepackage[most]{tcolorbox}
\usepackage{xcolor}
\newtcolorbox{highlighted}{colback=yellow,coltext=black,breakable}
\usepackage{xr}
\graphicspath{{./}}

\startlocaldefs
\newcommand{\CV}{\mbox{$\mathop{\mathtt{CV}}\limits$}\xspace}
\newcommand{\LOV}{\mbox{$\mathop{\mathtt{LOV}}\limits$}\xspace}
\newcommand{\minus}{\scalebox{0.6}{$-$}}
\newcommand{\plus}{\scalebox{0.6}{$+$}}
\newcommand{\BMTMKL}{\mbox{$\mathop{\mathtt{BMTMKL}}\limits$}\xspace}
\newcommand{\MTCRP}{\mbox{$\mathop{\mathtt{PLTR}}\limits$}\xspace}
\newcommand{\MTCRPH}{\mbox{$\mathop{\mathtt{PLTR_h}}\limits$}\xspace}
\newcommand{\KRL}{\mbox{$\mathop{\mathtt{KRL}}\limits$}\xspace}
\newcommand{\LETOR}{\mbox{$\mathop{\mathtt{LETOR}}\limits$}\xspace}

\newcommand{\feature}{\mbox{$\mathop{f}\limits$}\xspace}
\newcommand{\patient}{\mbox{$\mathop{\mathcal{P}}\limits$}\xspace}
\endlocaldefs

\begin{document}

\begin{frontmatter}

\begin{fmbox}
\dochead{Research}


\title{Cognitive Biomarker Prioritization in Alzheimer's 
Disease using Brain Morphometric Data}

\author[
   addressref={aff1},                   
   email={peng.707@buckeyemail.osu.edu} 
]{\inits{BP}\fnm{Bo} \snm{Peng}}
\author[
   addressref={aff2},
   email={Xiaohui.Yao@pennmedicine.upenn.edu}
]{\inits{XY}\fnm{Xiaohui} \snm{Yao}}
\author[
   addressref={aff3},
   email={srisache@iupui.edu}
]{\inits{SL}\fnm{Shannon L.} \snm{Risacher}}
\author[
   addressref={aff3},
   email={asaykin@iupui.edu}
]{\inits{AJ}\fnm{Andrew J.} \snm{Saykin}}
\author[
   addressref={aff2},
   email={li.shen@pennmedicine.upenn.edu}
]{\inits{LS}\fnm{Li} \snm{Shen}}
\author[
   addressref={aff1},
   corref={aff1},
   email={ning.104@osu.edu}
]{\inits{XN}\fnm{Xia} \snm{Ning}}
\author[
  noteref={n2}
]{\inits{ADNI}\fnm{for the} \snm{ADNI}}


\address[id=aff1]{
  \orgname{The Ohio State University}, 
  \city{Columbus},                                  
  \cny{US}                                              
}
\address[id=aff2]{%
  \orgname{University of Pennsylvania},
  \city{Philadelphia},
  \cny{US}
}
\address[id=aff3]{%
  \orgname{Indiana University},
  \city{Indianapolis},
  \cny{US}
}


\begin{artnotes}
%
\note[id=n2]{Data used in preparation of this article were obtained from the Alzheimer’s Disease Neuroimaging 
Initiative (ADNI) database (adni.loni.usc.edu). As such, the investigators within the ADNI contributed to the design and 
implementation of ADNI and/or provided data but did not participate in analysis or writing of this report. A complete listing of 
ADNI investigators can be found at: https://adni.loni.usc.edu/wp-content/uploads/how\_to\_apply/ADNI\_Data\_Use\_Agreement.pdf. \\
This work has been accepted by BMC MIDM. Copyright
may be transferred without notice, after which this version may no longer
be accessible.
}
\end{artnotes}



\begin{abstractbox}

\begin{abstract} 
\parttitle{Background}
Cognitive assessments represent the most common clinical routine for the diagnosis of Alzheimer's 
{Disease} 
(AD). Given a large
number of cognitive assessment tools and time-limited office visits, 
it is important to determine a proper set of cognitive tests for different subjects. 
%
Most current studies create guidelines of cognitive test selection for a targeted population, but they are not customized for each individual subject.
In this manuscript, we develop a machine learning paradigm enabling personalized cognitive assessments prioritization.

\parttitle{Method}
We adapt a newly developed learning-to-rank approach \MTCRP to implement our paradigm. This method learns the latent 
scoring function that pushes the most {effective} cognitive assessments onto the top of the prioritization list. 
We also extend \MTCRP to better separate the most effective cognitive assessments and the less effective ones.

\parttitle{Results}
Our empirical study on the ADNI data shows that the proposed paradigm outperforms 
{the state-of-the-art} 
baselines on identifying and prioritizing individual-specific cognitive biomarkers.
We conduct experiments in cross validation and level-out validation settings. 
In the two settings, 
our paradigm {significantly} outperforms the best baselines with improvement as much 
as 22.1\% and 19.7\%, respectively, on prioritizing cognitive features.

\parttitle{Conclusions}
The proposed paradigm achieves superior performance  
on prioritizing cognitive biomarkers. The cognitive biomarkers prioritized on top
have great potentials to facilitate personalized diagnosis,  
disease subtyping, and ultimately precision medicine in AD.
\end{abstract}



\end{abstractbox}
\end{fmbox}

\end{frontmatter}

\section{Background}
\label{sec:intro}

Identifying structural brain changes related to cognitive impairments is an important research topic in Alzheimer's Disease (AD) study. 
Regression models have been extensively studied to predict cognitive outcomes using morphometric measures 
that are extracted from structural magnetic resonance imaging (MRI) scans \cite{wan2014, yan2015}.
These studies are able to advance our understanding on the neuroanatomical basis of cognitive impairments. 
However, they are not designed to have direct impacts on clinical practice. 
To bridge this gap, in this manuscript we develop a novel learning paradigm to rank cognitive assessments 
based on their relevance to AD using brain MRI data.

Cognitive assessments represent the most common clinical routine for AD diagnosis. 
Given a large number of cognitive assessment tools and time-limited office visits, 
it is important to determine a proper set of cognitive tests for the subjects. 
Most current studies 
create guidelines of cognitive test selection for a targeted population \cite{Cordell2013, Scott2018}, 
but they are not customized for each individual subject.
In this work, we develop a novel learning paradigm that incorporate the ideas of precision medicine 
and customizes the cognitive test selection process to the characteristics of each individual patient. 
Specifically, we conduct a novel application of a newly developed learning-to-rank approach, 
denoted as \MTCRP \cite{he2018tcbb}, to the structural MRI and cognitive assessment data of 
the Alzheimer's Disease Neuroimaging Initiative (ADNI) cohort \cite{Weiner17}. 
Using structural MRI measures as the individual characteristics, we are able to not only identify 
individual-specific cognitive biomarkers but also prioritize them and their corresponding assessment 
tasks according to AD-specific abnormality. 
{We also extend \MTCRP to \MTCRPH using hinge loss \cite{gentile1999linear} to} 
 {more effectively prioritize individual-specific cognitive biomarkers.}
The study presented in this manuscript is a substantial extension from our preliminary study~\cite{Peng2019}. 

Our study is unique and innovative from the following two perspectives. 
First, conventional regression-based studies for  
cognitive performance prediction using MRI data focus on identifying relevant imaging biomarkers at the population level. 
However, our proposed model aims to identify AD-relevant cognitive biomarkers customized to each individual patient. 
Second, the identified cognitive biomarkers and assessments are prioritized based on 
the individual's brain characteristics. Therefore, they can be used to guide the selection of 
cognitive assessments in a personalized manner in clinical practice; 
it has the potential to enable personalized diagnosis and disease subtyping.

\subsection{Literature Review}
\label{sec:literature}
%
\subsubsection{Learning to Rank}
\label{sec:literature:LR}
%
Learning-to-Rank (\LETOR)~\cite{Liu2011} is a popular technique used in 
information retrieval~\cite{Li2011}, 
web search~\cite{agichtein2006} and recommender systems~\cite{Karatzoglou2013}.
Existing \LETOR methods can be classified into three categories~\cite{Liu2011}.
The first category is point-wise methods~\cite{Cao2007}, in which a function is learned to score individual instance, 
and then instances are sorted/ranked based on their scores.
The second category is pair-wise methods~\cite{Burges2007}, which maximize the number of correctly ordered pairs 
in order to learn the optimal ranking structure among instances.
The last category is list-wise methods~\cite{Lebanon2002}, in which a ranking function is learned to explicitly 
model the entire ranking.
Generally, pairwise and listwise methods have superior performance over point-wise methods
due to their ability to leverage order structure among instances in learning~\cite{Liu2011}.
%
Recently, \LETOR has also been applied in drug discovery and drug selection~\cite{Liu2017,Zhang2015,Liu2017a,Liu2017b}.
For example, Agarwal \etal~\cite{Agarwal2010} developed a bipartite ranking method to prioritize drug-like compounds.
He \etal~\cite{he2018tcbb} developed a joint push 
and learning-to-rank method to select cancer drugs for each individual patient.
These studies demonstrate the great potential of \LETOR 
in computational biology and computational medicine, particularly for biomarker prioritization.
%

\subsubsection{Machine Learning for AD Biomarker Discovery}
\label{sec:literature:AD}

The importance of {using big data} to enhance AD biomarker study has been widely recognized \cite{Weiner17}. 
As a result, numerous data-driven machine learning models have been developed for early AD detection 
and  AD-relevant biomarker identification including cognitive measures. 
These models are often designed to accomplish tasks such as 
classification (e.g., \cite{wang2017ipmi}), 
regression (e.g., \cite{wan2014,yan2015,yan2019frontneurosci}) or 
both (e.g., \cite{wang2012bioinfo2,brand2018miccai}), 
where imaging and other biomarker data are used to predict diagnostic, cognitive and/or other outcome(s) of interest. 
%
A drawback of these methods is that, although outcome-relevant biomarkers can be identified, 
they are identified at the population level and not specific to any individual subject. To bridge this gap, 
we 
{adapt} 
the \MTCRP method  for biomarker prioritization at the individual level, 
which has greater potential to directly impact personalized diagnosis. 

\section{Methods}
\label{sec:Med}

\subsection{Materials}
\label{sec:materials}

The imaging and cognitive data used in our study were obtained from the Alzheimer's Disease Neuroimaging Initiative (ADNI) 
database  \cite{Weiner17}. 
The ADNI was launched in 2003 as a public-private partnership, led by Principal Investigator Michael W. Weiner, MD. 
The primary goal of ADNI has been to test whether serial MRI, PET, other biological markers, and clinical and 
neuropsychological assessment can be combined to measure the progression of mild cognitive impairment 
(MCI, a prodromal stage of AD) and early AD. For up-to-date information, Please refer to~\cite{adni} for more detailed, 
up-to-date information.

Participants include 819 ADNI-1 subjects with 229 healthy control (HC), 397 MCI and 193 AD participants. 
We consider both MCI and AD subjects as patients, and thus we have 590 cases and 229 controls.
We downloaded the 1.5T baseline MRI scans and cognitive assessment data from the ADNI website~\cite{adni}. 
We processed the MRI scans using Freesurfer version 5.1~\cite{Risacher2013}, 
where volumetric and cortical thickness measures of 101 regions relevant to AD were extracted to characterize 
brain morphometry. 

We focus our analysis on 151 scores assessed in 15 neuropsychological tests. For convenience, 
we denote these measures as {\em cognitive features} and these tests as {\em cognitive tasks}. 
The 15 studied tasks include Alzheimer's Disease Assessment Scale (ADAS), Clinical Dementia Rating Scale (CDR),
Functional Assessment Questionnaire (FAQ), Geriatric Depression Scale (GDS), Mini-Mental State Exam (MMSE), 
Modified Hachinski Scale (MODHACH), Neuropsychiatric Inventory Questionnaire (NPIQ), Boston Naming Test (BNT), 
Clock Drawing Test (CDT), Digit Span Test (DSPAN), Digit Symbol Test (DSYM), Category Fluency Test (FLUENCY), 
Weschler's Logical Memory Scale (LOGMEM), Rey Auditory Verbal Learning Test (RAVLT) and Trail Making Test (TRAIL).



\subsection{Joint Push and Learning-To-Rank using Scores -- \MTCRP}
\label{sec:method:PLTR}

%
We use the joint push and learning-to-rank method that we developed in He \etal~\cite{he2018tcbb},
denoted as \MTCRP, for personalized cognitive feature prioritization.
\MTCRP has also been successfully applied in our preliminary study~\cite{Peng2019}. 
We aim to prioritize cognitive features for each individual patient that are most relevant to
his/her disease diagnosis. We will 
{use} 
patients' brain morphometric measures that are extracted from their MRI scans for the 
cognitive feature prioritization. 
The cognitive features are in the form of scores or answers in the cognitive tasks that the patients take. 
The prioritization outcomes can potentially be used in clinical practice to suggest the most relevant 
cognitive features or tasks that can most effectively facilitate diagnosis of an individual subject. 
%
%
%

In order to prioritize MCI/AD cognitive features,  
\MTCRP learns and uses patient latent vector representations and their imaging features to
score each cognitive feature for each individual patient. Then, \MTCRP ranks the cognitive features based on their scores.  
Patients with similar imaging feature profiles will have similar latent vectors and thus similiar ranking of cognitive features {\cite{cf,wang2006unifying}}. 
During the learning, \MTCRP explicitly pushes the most relevant cognitive features on
top of the less relevant features for each patient, and therefore
optimizes the latent patient vectors and cognitive feature vectors in a way that they will reproduce the feature ranking
structures {\cite{Liu2011}}. 
%
%
%
%
%
In \MTCRP, these latent vectors are learned via solving the following optimization problem:
%
\begin{equation}
  \label{eqn:obj}
    \min_{U, V} {\mathcal{L}_s} = (1 - \alpha) {P^{\uparrow}_s} + \alpha {O^{\plus}_s}
    + \frac{\beta}{2} R_{uv} + \frac{\gamma}{2} R_{\text{csim}},
\end{equation}
where
$\alpha$, $\beta$ and $\gamma \in [0,1]$ are coefficients of $O^{\plus}_s$,
$R_{uv}$ and $R_{\text{csim}}$ terms, respectively;
\mbox{$U = [\mathbf{u}_1, \mathbf{u}_2, \cdots, \mathbf{u}_m]$}
and \mbox{$V=[\mathbf{v}_1, \mathbf{v}_2, \cdots, \mathbf{v}_n]$}
are the latent matrices for patients and features, respectively ($\mathbf{u}$ and $\mathbf{v}$ 
are column latent patient vector and feature vector, respectively); 
${\mathcal{L}_s}$ is the overall loss function. In Problem~\ref{eqn:obj}, 
%
${P^{\uparrow}_s}$ measures the average number of relevant cognitive features 
ranked below an irrelevant cognitive feature, defined as follows, 
%
\begin{equation}
  \label{eqn:push}
  {P^{\uparrow}_s} = \sum\limits_{p = 1}^m
  \frac{1}{n_p^{\plus}n_p^{\minus}}\sum\limits_{\scriptsize{\feature^{\minus}_i\in \patient_p}}
  \sum\limits_{\scriptsize{\feature_j^{\plus}\in \patient_p^{\plus}}}
  \mathbb{I} ( {s_p}(\feature_j^{\plus}) \le {s_p}(\feature_i^{\minus}) ),
\end{equation}
where $m$ is the number of patients, $\feature^{\plus}_j$ and $\feature^{\minus}_i$ are the
relevant and irrelevant features of patient $\patient_p$, 
$n^{\plus}_p$ and $n^{\minus}_p$ are their respective numbers, and
$\mathbb{I}(x)$ is the indicator function 
($\mathbb{I}(x) = 1$ if $x$ is true, otherwise 0).
In Equation~(\ref{eqn:push}), $s_p(\feature_i)$ is a scoring function defined as follows, 
%
\begin{ceqn}
  \label{eqn:pred}
  \begin{align}
    s_p(\feature_i) = \mathbf{u}^{\mathsf{T}}_p \mathbf{v}_i, 
  \end{align}
\end{ceqn}
that is, it calculates the score of feature $\feature_i$ on patient $\patient_p$ using their respective latent
vectors $\mathbf{u}_p$ and $\mathbf{v}_i${\cite{koren2009matrix}}. 
By minimizing $P^{\uparrow}_s$, \MTCRP learns to assign higher scores to relevant features than irrelevant features
so as to rank the relevant features at the top of the final ranking list.
Note that, \MTCRP learns different latent vectors and ranking lists for different subjects, and therefore enables
personalized feature prioritization.
In Problem~(\ref{eqn:obj}), ${O^{\plus}_s}$ measures the ratio of mis-ordered feature pairs over the relevant features among
all the subjects, defined as follows,
%
\begin{eqnarray}
  \begin{aligned}
  \label{eqn:order_rel}
  {O^{\plus}_s}  
     	 = \sum\limits_{p=1}^m \frac{1}{|\{\feature_i^{\plus} \succ_{\scriptsize{\patient_p}} \feature_j^{\plus}\}|}
         \sum\limits_{\scriptsize{\feature_i^{\plus} \succ_{\tiny{\patient_p}} \feature_j^{\plus}}}
         \mathbb{I}({s_p}(\feature_i^{\plus}) < {s_p}(\feature_j^{\plus})),
         \hspace{7pt} 
  \end{aligned}
  %
\end{eqnarray}
where $\feature_i \succ_{\scriptsize{\patient_p}} \feature_j$ represents
that $\feature_i$ is ranked higher than $\feature_j$ for patient $\patient_p$.
By minimizing $O^{\uparrow}_s$, \MTCRP learns to push the most relevant features on top of the less 
relevant features. Thus, most relevant features are pushed to the very top of the ranking list.
In Problem~\eqref{eqn:obj}, $R_{uv}$ is a regularizer on $U$ and $V$ to prevent overfitting,
defined as, 
\begin{equation}
  \label{eqn:reg}
    R_{uv} = \frac{1}{m}\|U\|^2_{{F}} + \frac{1}{n}\|V\|^2_{{F}}, 
\end{equation}
where $\|X\|_{{F}}$ is the Frobenius norm of matrix $X$.
%
%
%
$R_{\text{csim}}$ is a regularizer on patients
to constrain patient latent vectors, defined as
%
\begin{equation}
  \label{eqn:sim}
    R_{\text{csim}} = \frac{1}{m^2}\sum_{p=1}^m\sum_{q=1}^m w_{pq}\|\mathbf{u}_p-\mathbf{u}_q\|_2^2,    
\end{equation}
where $w_{pq}$ is the similarity between subject $\patient_p$ and $\patient_q$ that is calculated
using the imaging features of the subjects. 
The assumption here is that patients who are similar in terms of imaging features 
could also be similar in terms of cognitive features.
%
%

\subsection{Joint Push and Learning-To-Rank with Marginalization -- \MTCRPH}
\label{sec:method:PLTRH}
The objective of \MTCRP is to score 
relevant features higher than less relevant features 
as shown in Equation~\ref{eqn:push} and Equation~\ref{eqn:order_rel}.
However, in some cases, the score of relevant features is expected to be higher than that of less relevant features 
by a large margin.
For example, patients can be very sensitive to a few cognitive tasks but less sensitive to many others.
%
In order to incorporate such information, we propose a {new}
hinge loss~\cite{gentile1999linear}
based \MTCRP, denoted as \MTCRPH.
In \MTCRPH, the overall loss function is very similar to Equation~\ref{eqn:obj}, defined as follows,
\begin{equation}
  \label{eqn:H:obj}
    \min_{U, V} {\mathcal{L}_h} = (1 - \alpha) {P^{\uparrow}_h} + \alpha {O^{\plus}_h}
    + \frac{\beta}{2} R_{uv} + \frac{\gamma}{2} R_{\text{csim}},
\end{equation}
where ${\mathcal{L}_h}$ is the overall loss function; 
$U$, $V$, $R_{uv}$ and $R_{\text{csim}}$ are identical as those in Equation~\ref{eqn:obj}. 
In \MTCRPH, ${P^{\uparrow}_h}$ measures the average loss between the relevant features
 and irrelevant features using hinge loss as follows,
\begin{eqnarray}
  \begin{aligned}
  \label{eqn:H:push}
  {P^{\uparrow}_h} = \sum\limits_{\mathclap{p = 1}}^m
  \frac{1}{n_p^{\plus}n_p^{\minus}}\!\!\sum\limits_{{\scriptsize{\feature^{\minus}_i\in \patient_p}}}
  \sum\limits_{\mathclap{\scriptsize{~~\feature_j^{\plus}\in \patient_p^{\plus}}}}
  \text{max} (0, t_p\hspace{-0.3em} -\hspace{-0.3em} ( {s_p}(\feature_j^{\plus})\hspace{-0.3em} -\hspace{-0.3em} {s_p}(\feature_i^{\minus}) ) ),
  \hspace{7pt}
  \end{aligned}
\end{eqnarray}
where $\text{max} (0, t_p - ( {s_p}(\feature_j^{\plus}) - {s_p}(\feature_i^{\minus}) ) )$ is the hinge loss
($\text{max}(0, x) = x \text{ if $x>0$}$, otherwise $0$)
between the relevant feature $\feature_j^{\plus}$ and the irrelevant feature $\feature_i^{\minus}$, 
and $t_p$ is the pre-defined margin. 
%
%
Specifically, only when ${s_p}(\feature_j^{\plus}) - {s_p}(\feature_i^{\minus}) > t_p$ will 
not induce any loss during optimization. 
Otherwise, the hinge loss will be positive and increase as ${s_p}(\feature_j^{\plus}) - {s_p}(\feature_i^{\minus}) $ gets smaller
than $t_p$.
Thus, the hinge loss forces the scores of relevant features higher than those of irrelevant features by at least $t_p$. 
By doing this,  the relevant features are ranked higher than irrelevant features in the ranking list.
Similarly, ${O^{\plus}_h}$ measures the average loss among the relevant features 
also using hinge loss as follows,
\begin{eqnarray}
  \begin{aligned}
  \label{eqn:H:order_rel}
  {O^{\plus}_h}  
         = \sum\limits_{p=1}^m \frac{1}{|\{\feature_i^{\plus}\!\!\succ_{\mathclap{\scriptsize{\patient_p}}} \feature_j^{\plus}\}|}
         \sum\limits_{\mathclap{~~~\scriptsize{\feature_i^{\plus} \succ_{\tiny{\patient_p}} \feature_j^{\plus}}}}
	 \text{max}  (0, 
	 t_o \hspace{-0.3em}-\hspace{-0.3em} ( {s_p}(\feature_i^{\plus}) \hspace{-0.3em} - \hspace{-0.3em}{s_p}(\feature_j^{\minus}) ) ),
	 \hspace{7pt}
  \end{aligned}
  %
\end{eqnarray}
where 
$t_o$ is also the pre-defined margin.

\subsection{Data Processing}
\label{sec:data}

\subsubsection{Data Normalization}
\label{sec:materials:norm}

%
%
Following the protocol in our preliminary study~\cite{Peng2019}, 
we selected all the MCI and AD patients from ADNI
and conducted the following data normalization for these patients.
We first performed a $t$-test on each cognitive feature between patients and controls,
and selected those features if there is a significant difference
between patients and controls on these features.
%
Then, we converted the selected features into $[0, 1]$ by shifting and scaling the feature
values.
%
We also converted all the normalized feature values according to the Cohen's $d$ of the features between patients and controls,
and thus, smaller values always indicate higher AD possibility. 
After that, we filtered out features with values $0$, $1$ or $0.5$ for more than 95\% patients. 
This is to discard features that are either not discriminative, or extremely dominated by patients or controls.
{After the filtering step, we have 112 cognitive features remained and used in experiments. Table S1 in the supplementary 
materials presents these 112 cognitive features. }
We conducted the same process as above on the imaging features. {Table S2 in the supplementary materials presents 
these imaging features used in experiments. }

\subsubsection{Patient Similarities from Imaging Features}
\label{sec:data:sim}

Through the normalization and filtering steps as in Section~\ref{sec:materials:norm},
we have 86 normalized imaging features remained. We represent each patient using a vector of these features, 
denoted as $\mathbf{r}_p = [r_{p1}, r_{p2}, \cdots, r_{p86}]$, in which $r_{pi}$ ($i = 1, \cdots, 86$)
is an imaging feature for patient $p$. 
%
%
%
%
%
%
%
%
%
%
We calculate the patient similarity from imaging features using the radial basis function (RBF) kernel, that is, $w_{pq} = \exp( -\frac{\|\mathbf{r}_p - \mathbf{r}_q\|^2}{2\sigma^2})$,
%
%
where $w_{pq}$ is the patient similarity used in $R_{\text{csim}}$. 

\section{Results}
\label{sec:results}


\subsection{Baseline Methods}
\label{sec:method:baselines}
We compare \MTCRP and \MTCRPH with two baseline methods:
the Bayesian Multi-Task Multi-Kernel Learning (\BMTMKL) method~\cite{costello2014community}
and the Kernelized Rank Learning (\KRL) method~\cite{he2018kernelized}.

\subsubsection{Bayesian Multi-Task Multi-Kernel Learning (\BMTMKL)}
\label{sec:method:baselines:BMTBKL}
\BMTMKL is a {state-of-the-art} baseline for biomarker prioritization.
It was originally proposed to rank cell lines for drugs and won the DREAM 7
challenge~\cite{dream}.
%
In our study, \BMTMKL uses the multi-task and multi-kernel learning within kernelized regression to predict cognitive feature values
and learns parameters by conducting Bayesian inference.
We use the patient similarity matrix calculated from FreeSurfer features as the kernels in \BMTMKL.

\subsubsection{Kernelized Rank Learning (\KRL)}
\label{sec:method:baselines:KRL}
KRL represents another {state-of-the-art} baseline for biomarker prioritization.
In our study, \KRL uses kernelized regression with a ranking loss to learn the ranking structure of patients
and to predict the cognitive feature values.
The objective of \KRL is to maximize the hits among the top k of the ranking list.
%
We use the patient similarity matrix calculated from FreeSurfer features as the kernels in \KRL.
%

\subsection{Training-Testing Data Splits}
\label{sec:exp:split}

Following the protocol in our preliminary study~\cite{Peng2019}, 
we test our methods in two different settings: cross validation (\CV) and leave-out validation (\LOV).
In \CV, we randomly split each patient's cognitive tasks into
5 folds: all the features of a cognitive task will be either split into training or testing set.
We use 4 folds for training and the rest fold for testing, and do such experiments
5 times, each with one of the 5 folds as the testing set. The overall performance of the methods
is averaged over the 5 testing sets. This setting corresponds to the goal to prioritize additional cognitive tasks
that a patient should complete.
%
%
%
%
In \LOV, 
we split patients (not patient tasks) into training and testing sets, and a certain patient and all his/her
cognitive features will be either in the training set or in the testing set.
%
This corresponds to the use scenario to identify the most relevant cognitive tasks that
a new patient needs to take, based on the existing imaging information of the patient, when the patient
has not completed any cognitive tasks. 
Figure~\ref{fig:cv} and Figure~\ref{fig:5fold} demonstrate the \CV and \LOV data split processes, respectively.

Please note that as presented in Section~\ref{sec:materials:norm}, for 
normalized cognitive features, smaller values always 
indicate more AD possibility. Thus, in both settings,
we use the ranking list of normalized cognitive
features of each patient as ground truth for training and testing.

\begin{figure}[t!]
  \centering
    \begin{center}
	\definecolor{train}{HTML}{A1A1FF}
\definecolor{test}{HTML}{E2E3FF}

\begin{tikzpicture}[fill=white]

    \draw
    (0.0, 0.0) node {};

    \draw
    [fill=train] (0.6, 0.0) rectangle (1.1, 0.3) 
    (2.6, 0.15) node [text=black] {training patients};

    \draw
    [fill=test] (4.3, 0.0) rectangle (4.8, 0.3)
    (6.3, 0.15) node [text=black] {testing patients};

\end{tikzpicture}
        \scalebox{0.8}{
        \input{random_5_fold.tex}
      }
      \caption{Data split for cross validation (\CV)}
      \label{fig:cv}
      \end{center}
\end{figure}
%
  %
\begin{figure}[!t]
  \centering
  \begin{center}
      \definecolor{train}{HTML}{A1A1FF}
\definecolor{test}{HTML}{E2E3FF}

\begin{tikzpicture}[fill=white]

    \draw
    (0.0, 0.0) node {};

    \draw
    [fill=train] (0.6, 0.0) rectangle (1.1, 0.3) 
    (2.6, 0.15) node [text=black] {training patients};

    \draw
    [fill=test] (4.3, 0.0) rectangle (4.8, 0.3)
    (6.3, 0.15) node [text=black] {testing patients};

\end{tikzpicture}
      \scalebox{0.8}{
      \input{random_loo.tex}
    }
    \caption{Data split for leave-out validation (\LOV)}
    \label{fig:5fold}
  \end{center}
  %
\end{figure}

\subsection{Parameters}
\label{sec:method:parameters}


We conduct  grid search 
to identify the best parameters {on each evaluation metric}
for each model. 
We use 0.3 and 0.1 as the value of $t_p$ and $t_o$, respectively.
%
In the experimental results, we report the combinations of parameters 
that achieve the best performance on evaluation metrics.
{We implement {\MTCRP} and {\MTCRPH} using Python 3.7.3
and Numpy 1.16.2, and run the experiments on Xeon
E5-2680 v4 with 128G memory.}

\subsection{Evaluation Metrics}
\label{sec:method:metrics}

\begin{table*}[t]
  \vspace{-5pt}
  \centering
  \caption{Overall Performance in \CV}
  \label{table:cv}
  \begin{small}
  \begin{threeparttable}
      \begin{tabular}[]{
          @{\hspace{2pt}}l@{\hspace{5pt}}
	  @{\hspace{5pt}}c@{\hspace{5pt}}
          @{\hspace{5pt}}c@{\hspace{5pt}}
          @{\hspace{5pt}}c@{\hspace{5pt}}
          @{\hspace{5pt}}c@{\hspace{5pt}}
          @{\hspace{2pt}}c@{\hspace{2pt}}
          @{\hspace{5pt}}c@{\hspace{5pt}}
          @{\hspace{5pt}}c@{\hspace{5pt}}
          @{\hspace{5pt}}c@{\hspace{5pt}}
          @{\hspace{5pt}}c@{\hspace{5pt}}
          @{\hspace{5pt}}c@{\hspace{2pt}}       
        }
	\toprule
        \multirow{2}{*}{method} & \multicolumn{2}{c}{parameters} & \multicolumn{2}{c}{feature level} && \multicolumn{5}{c}{task level} \\
	\cmidrule{2-3} \cmidrule{4-5} \cmidrule{7-11}
	& d & $\lambda$  & QH@5  & WQH@5 && NH$_1@1$ & NH$_{2}@1$ & NH$_{3}@1$ & NH$_{5}@1$ & NH$_{\text{all}}@1$\\
        \midrule
	\multirow{6}{*}{\MTCRP}
        &10 & - & \underline{\textbf{2.665}}{$\pm$0.07} & 3.136{$\pm$0.12} && 0.605{$\pm$0.03} & 0.701{$\pm$0.04} & 0.713{$\pm$0.05} & 0.725{$\pm$0.05} & 0.683{$\pm$0.04} \\
        &10 & - & 2.647{$\pm$0.08} & \underline{\textbf{3.191}}{$\pm$0.14} && 0.599{$\pm$0.03} & 0.677{$\pm$0.04} & 0.707{$\pm$0.04} & 0.725{$\pm$0.05} & 0.677{$\pm$0.04} \\
        &10 & - & 2.569{$\pm$0.08} & 2.957{$\pm$0.11} && \textbf{0.635}{$\pm$0.03} & 0.707{$\pm$0.04} & 0.689{$\pm$0.05} & 0.719{$\pm$0.04} & 0.653{$\pm$0.04} \\
        &10 & - & 2.623{$\pm$0.06} & 3.073{$\pm$0.09} && 0.623{$\pm$0.03} & \underline{\textbf{0.713}}{$\pm$0.05} & 0.707{$\pm$0.04} & 0.719{$\pm$0.04} & 0.671{$\pm$0.04} \\
        &50 & - & 2.467{$\pm$0.07} & 2.992{$\pm$0.11} && 0.605{$\pm$0.03} & 0.695{$\pm$0.04} & \textbf{0.725}{$\pm$0.06} & 0.725{$\pm$0.04} & 0.653{$\pm$0.04} \\
        &30 & - & 2.491{$\pm$0.07} & 3.080{$\pm$0.14} && 0.563{$\pm$0.04} & 0.689{$\pm$0.05} & 0.713{$\pm$0.04} & \textbf{0.749}{$\pm$0.04} & \textbf{0.689}{$\pm$0.03} \\
	\midrule
	\multirow{6}{*}{\MTCRPH}
	& 10 & - & \textbf{2.599}{$\pm$0.09} & 3.111{$\pm$0.12} && 0.623{$\pm$0.02} & 0.671{$\pm$0.03} & 0.713{$\pm$0.03} & 0.719{$\pm$0.04} & \textbf{0.707}{$\pm$0.03} \\
	& 10 & - & 2.575{$\pm$0.08} & \textbf{3.115}{$\pm$0.13} && 0.623{$\pm$0.03} & 0.677{$\pm$0.03} & 0.737{$\pm$0.04} & 0.749{$\pm$0.03} & 0.695{$\pm$0.03} \\
	& 10 & - & 2.419{$\pm$0.09} & 2.827{$\pm$0.12} && \underline{\textbf{0.647}}{$\pm$0.03} & 0.695{$\pm$0.03} & 0.671{$\pm$0.03} & 0.707{$\pm$0.03} & 0.635{$\pm$0.03} \\
	& 30 & - & 2.138{$\pm$0.10} & 2.583{$\pm$0.18} && 0.629{$\pm$0.02} & \textbf{0.701}{$\pm$0.02} & 0.695{$\pm$0.03} & 0.695{$\pm$0.04} & 0.593{$\pm$0.05} \\
	& 50 & - & 2.102{$\pm$0.07} & 2.470{$\pm$0.10} && 0.533{$\pm$0.03} & 0.677{$\pm$0.03} & \underline{\textbf{0.743}}{$\pm$0.04} & 0.754{$\pm$0.03} & 0.629{$\pm$0.05} \\
	& 30 & - & 2.281{$\pm$0.07} & 2.768{$\pm$0.18} && 0.563{$\pm$0.03} & 0.689{$\pm$0.03} & 0.707{$\pm$0.04} & \textbf{0.760}{$\pm$0.05} & 0.701{$\pm$0.05} \\
	\midrule
	\multirow{2}{*}{\KRL}
	& - & 2 & \textbf{2.102}{$\pm$0.26} & \textbf{2.167}{$\pm$0.37} && \textbf{0.569}{$\pm$0.03} & \textbf{0.611}{$\pm$0.05} & \textbf{0.635}{$\pm$0.04} & \textbf{0.683}{$\pm$0.03} & 0.689{$\pm$0.07} \\
	& - & 1.5 & 2.078{$\pm$0.15} & 2.143{$\pm$0.25} && 0.503{$\pm$0.04} & 0.575{$\pm$0.05} & 0.617{$\pm$0.05} & 0.677{$\pm$0.04} & \underline{\textbf{0.760}}{$\pm$0.06} \\
	\midrule
	\BMTMKL
	& - & - & \textbf{2.443}{$\pm$0.12} & \textbf{2.614}{$\pm$0.20} && \textbf{0.413}{$\pm$0.07} & \textbf{0.491}{$\pm$0.08} & \textbf{0.593}{$\pm$0.05} & \underline{\textbf{0.784}}{$\pm$0.05} & \textbf{0.749}{$\pm$0.05} \\
        \bottomrule
      \end{tabular}
      \begin{tablenotes}
        \setlength\labelsep{0pt}
	\begin{scriptsize}
	\item
          The column ``$d$'' corresponds to the latent
        dimension. 
        The numbers in the form of $x{\pm}y$ represent the mean ($x$) and standard deviation ($y$). 
        The best performance of each method is in
        \textbf{bold}. 
	The best performance under each evaluation metric is \underline{underlined}. \par
        \end{scriptsize}
      \end{tablenotes}
  \end{threeparttable}
  \end{small}
\end{table*}


\subsubsection{Metrics on Cognitive Feature Level}
\label{sec:methods:metrics:question}

We use a metric named average feature hit at $k$ (QH@$k$) as in our preliminary study~\cite{Peng2019} to evaluate the ranking performance, 
%
\begin{equation}
  \label{eqn:qh}
  \text{QH}@k(\mathbf{\tau}^q, \tilde{\mathbf{\tau}}^q) = \sum\nolimits_{i = 1}^k \mathbb{I}(\tilde{{\tau}}^q_i \in {\mathbf{\tau}^q(1:k)}),
\end{equation}
where $\mathbf{\tau}^q$ is the ground-truth ranking list of all the features in all the tasks,
$\mathbf{\tau}^q(1:k)$ is the top $k$ features in the list, 
$\tilde{\mathbf{\tau}}^q$ is the predicted ranking list of all the features, and 
$\tilde{\mathbf{\tau}}^q_i$ is the $i$-th ranked features in $\tilde{\mathbf{\tau}}^q$. 
That is, QH@$k$ calculates the number of features among top $k$ in the predicted feature lists that are also in
the ground truth (i.e., hits).
Higher QH@$k$ values indicate better prioritization performance.  

We use a second evaluation metric weighted average feature hit at $k$ (WQH@$k$) as follows:
%
\begin{equation}
  \label{eqn:wqh}
  \text{WQH}@k(\mathbf{\tau}^q, \tilde{\mathbf{\tau}}^q) = {\sum\nolimits_{j=1}^k  QH@j(\mathbf{\tau}^q, \tilde{\mathbf{\tau}}^q)}/k,
\end{equation}
that is, $\text{WQH}@k$ is a weighted version of $\text{QH}@k$ that calculates the average of $\text{QH}@j$ ($j = 1, \cdots, k$)
over top $k$.
Higher $\text{WQH}@k$ indicates more feature hits and those hits are ranked on top in the ranking list.
%

\subsubsection{Metrics on Cognitive Task Level}
\label{sec:methods:metrics:questionnaire}


In in Peng \etal~\cite{Peng2019}, 
we use the mean of the top-$g$ normalized ground-truth scores/predicted scores on the features of each
cognitive task for a patient as the score of that task for that patient.
%
For each patient, we rank the tasks using their ground-truth scores and use the ranking
as the ground-truth ranking of these tasks.
Thus, these scores measure how much relevant to AD the task indicates for the patients.
We use the predicted scores to rank cognitive tasks into the predicted ranking of the tasks.
We define a third evaluation metric task hit at $k$ (NH$_g$@$k$) as follows  
to evaluate the ranking performance in terms of tasks, 
%
\begin{equation}
  \label{eqn:nh}
  \text{NH}_g@k(\mathbf{\tau}_g^n, \tilde{\mathbf{\tau}}_g^n) = \sum\nolimits_{i = 1}^k \mathbb{I}(\tilde{{\tau}}^n_{gi} \in {\mathbf{\tau}_g^n(1:k)}),
\end{equation}
where $\mathbf{\tau}_g^n$/$\tilde{\mathbf{\tau}}_g^n$ is the ground-truth/predicted
ranking list of all the tasks using top-$g$ question scores. 
%
%
%
%
%

\section{Experimental Results}
\label{sec:results}

\subsection{Overall Performance on \CV}
\label{sec:results:overall:cv}

Table~\ref{table:cv} presents the performance of \MTCRP, \MTCRPH and two baseline methods in the \CV setting.
{Note that overall, {\MTCRP} and {\MTCRPH} have similar standard deviations; {\KRL} and {\BMTMKL}
have higher standard deviations compared to {\MTCRP} and {\MTCRPH}. This indicates that {\MTCRP} and {\MTCRPH}
are more robust than {\KRL} and {\BMTMKL} for the prioritization tasks. }

%
%
%

%
\subsubsection{Comparison on cognitive feature level}
\label{sec:results:overall:cv:feature}
%
For cognitive features from all tasks,
\MTCRP is able to identify on average 2.665{{$\pm$}0.07} out of the top-5 most relevant ground-truth cognitive features
 among its top-5 predictions (i.e., QH@5=2.665{$\pm$0.07}).
\MTCRPH achieves similar performance as \MTCRP, 
and identifies on average 2.599{$\pm$0.09} most relevant ground-truth cognitive features on its top-5 predictions (i.e., QH@5=2.599{$\pm$0.09}).
\MTCRP and \MTCRPH significantly
outperform the baseline methods in terms of all the evaluation metrics on cognitive feature level (i.e., QH@5 and WQH@5).
Specifically, \MTCRP outperforms the best baseline method \BMTMKL at 9.1{$\pm$3.7}\% and 22.1{$\pm$9.5}\% 
on QH@5 and WQH@5, respectively.
\MTCRPH also outperforms \BMTMKL at 6.4{$\pm$4.3}\% and 19.2{$\pm$10.1}\% on QH@5 and WQH@5, respectively.
These experimental results demonstrate that among the top 5 features in the ranking list, 
\MTCRP and \MTCRPH are able to rank more relevant features on top than 
the two {state-of-the-art}
 baseline methods and the positions of those hits are also higher than those in the baseline methods.
%

\subsubsection{Comparison on cognitive task level}
\label{sec:results:overall:cv:task}
%
For the scenario to prioritize cognitive tasks that each patient
should take, \MTCRP and \MTCRPH are able to identify the top-1 most relevant task for 72.5{$\pm$6.0}\% and 74.3{$\pm$4.0}\% of all the
patients when using 3 features to score cognitive tasks, respectively (i.e., NH$_3$=0.725{$\pm$0.06} for \MTCRP and NH$_3$=0.743{$\pm$0.04} for 
\MTCRPH). 
This indicates the strong power of \MTCRP and \MTCRPH in prioritizing cognitive features
and in recommending relevant cognition tasks for real clinical applications. 
We also find that \MTCRP and \MTCRPH are able to outperform baseline methods on most of the metrics 
on cognitive task level (i.e., NH$_g@1$).
\MTCRP outperforms the best baseline method at 11.6{$\pm$5.6}\%, 16.7{$\pm$6.1}\% and 14.2{$\pm$6.6}\% on NH$_1@1$, 
NH$_2@1$ and NH$_3@1$, respectively.
\MTCRPH performs even better than \MTCRP on NH$_1@1$ and NH$_3@1$, in addition to that 
it outperforms the best performance of baseline methods at 13.7{$\pm$5.3}\%, 14.7{$\pm$4.8}\% and 17.0{$\pm$8.8}\% on NH$_1@1$, 
NH$_2@1$ and NH$_3@1$, respectively.
\MTCRP and \MTCRPH perform slightly worse than baseline methods on NH$_5@1$ and NH$_\text{all}@1$ 
(0.760{$\pm$0.05} vs 0.784{$\pm$0.05} on NH$_5@1$ and 0.707{$\pm$0.03} vs 0.760{$\pm$0.06} on NH$_\text{all}@1$).
These experimental results indicate that \MTCRP and \MTCRPH are able to push the most 
relevant task to the top of the ranking list than baseline methods when using a small number 
of features to score cognitive tasks.
%
%
Note that in \CV, each patient has only a few cognitive tasks in the testing set. Therefore, we only
consider the evaluation at the top task in the predicted task rankings (i.e.,
only NH$_g@1$ in Table~\ref{table:cv}).
%

%
Table~\ref{table:cv} also shows that \MTCRPH outperforms \MTCRP
on most of the metrics on cognitive task level (i.e., NH$_g@1$).
\MTCRPH outperforms \MTCRP at 1.9{$\pm$0.5}\%, 2.5{$\pm$1.2}\%, 1.5{$\pm$0.3}\% and 2.6{$\pm$0.9}\%
on NH$_1$@1, NH$_3$@1, NH$_5$@1 and NH$_\text{all}$@1, respectively.
This indicates that generally \MTCRPH is better than \MTCRP on ranking cognitive tasks in \CV setting.
The reason could be that the hinge-based loss functions with pre-defined margins can enable significant
difference between the scores of
relevant features and irrelevant features, and 
thus effectively push relevant features upon irrelevant features.

\subsection{Overall Performance on \LOV}
\label{sec:results:overall:lov}

\begin{table*}[t]
  \centering
  \caption{Overall Performance in \LOV on 26 testing patients} 
  \label{table:lov:26}
  \begin{small}
  \begin{threeparttable}
    \begin{tabular}[]{
        @{\hspace{2pt}}l@{\hspace{2pt}}
        %
        %
        @{\hspace{3pt}}c@{\hspace{3pt}}
	@{\hspace{2pt}}c@{\hspace{1pt}}
	@{\hspace{1pt}}c@{\hspace{1pt}}
        @{\hspace{3pt}}c@{\hspace{3pt}}
        @{\hspace{2pt}}c@{\hspace{5pt}}
        @{\hspace{3pt}}c@{\hspace{3pt}}
        @{\hspace{2pt}}c@{\hspace{5pt}}
        @{\hspace{3pt}}c@{\hspace{3pt}}
        @{\hspace{2pt}}c@{\hspace{5pt}}
        @{\hspace{3pt}}c@{\hspace{3pt}}
        @{\hspace{2pt}}c@{\hspace{5pt}}
        @{\hspace{3pt}}c@{\hspace{3pt}}
        @{\hspace{2pt}}c@{\hspace{5pt}}
        @{\hspace{3pt}}c@{\hspace{2pt}}
        }
	\toprule
	\multirow{2}{*}{method} 
	& \multicolumn{2}{c}{feature level} && \multicolumn{10}{c}{task level} \\
	\cmidrule{2-3} \cmidrule{5-14}
	& QH@5 & WQH@5
        && NH$_1$@1 & NH$_1$@5 & NH$_2$@1 & NH$_2$@5
        & NH$_3$@1 & NH$_3$@5 & NH$_5$@1 & NH$_5$@5
        & NH$_{\text{all}}$@1 & NH$_{\text{all}}$@5 \\
        \midrule
	\multirow{5}{*}{\MTCRP}
& \underline{\textbf{1.615}} & \underline{\textbf{1.906}} 
&& 0.846 & 3.231 & 0.577 & 3.385 & 0.231 & 3.654 & 0.308 & 3.346 & 0.808 & 3.692 \\
& 1.500 & 1.778 
&& \underline{\textbf{0.846}} & \textbf{3.269} & \underline{\textbf{0.577}} & \underline{\textbf{3.538}} & 0.269 & 3.654 & 0.269 & 3.269 & 0.808 & 3.577 \\
& 1.538 & 1.856 
&& 0.846 & 3.192 & 0.577 & 3.423 & \textbf{0.308} & \underline{\textbf{3.731}} & 0.346 & 3.346 & 0.808 & 3.615 \\
& 1.577 & 1.851 
&& 0.846 & 3.192 & 0.577 & 3.462 & 0.308 & 3.654 & \underline{\textbf{0.346}} & \textbf{3.462} & 0.808 & 3.654 \\
& 1.615 & 1.906 
&& 0.846 & 3.231 & 0.577 & 3.385 & 0.231 & 3.654 & 0.308 & 3.346 & \underline{\textbf{0.808}} & \textbf{3.692} \\
	\midrule
	\multirow{4}{*}{\MTCRPH}
& \underline{\textbf{1.615}} & 1.836
&& \underline{\textbf{0.846}} & 3.192 & \underline{\textbf{0.577}} & \textbf{3.500} & \textbf{0.269} & 3.731 & \underline{\textbf{0.346}} 
& 3.731 & \underline{\textbf{0.808}} & 4.154 \\
& 1.538 & \textbf{1.891}
&& 0.846 & 3.192 & 0.577 & 3.500 & 0.269 & 3.731 & 0.346 & 3.615 & 0.808 & 4.038 \\
& 1.538 & 1.856
&& 0.769 & \textbf{3.308} & 0.577 & 3.462 & 0.269 & 3.615 & 0.308 & 3.385 & 0.808 & 3.500 \\
& 1.538 & 1.712
&& 0.846 & 3.115 & 0.577 & 3.423 & 0.154 & \underline{\textbf{3.731}} & 0.308 & \textbf{3.808} & 0.808 & \underline{\textbf{4.269}} \\
	\midrule
	\multirow{3}{*}{\KRL}
& \textbf{1.423} & 1.656
&& 0.615 & 2.615 & \underline{\textbf{0.577}} & 3.308 & 0.038 & 3.577 & \underline{\textbf{0.346}} & \underline{\textbf{3.962}} 
& \underline{\textbf{0.808}} & \underline{\textbf{4.269}} \\
&  1.346 & \textbf{1.881}
&& 0.577 & 2.615 & 0.577 & 3.308 & 0.038 & 3.577 & 0.346 & 3.962 & 0.808 & 4.269 \\
& 1.346 & 1.435
&& \textbf{0.808} & \underline{\textbf{3.423}} & 0.538 & \textbf{3.500} & \underline{\textbf{0.346}} & \underline{\textbf{3.731}} 
& 0.154 & 3.423 & 0.808 & 3.538 \\
	\midrule
	\BMTMKL
& \textbf{0.423} & \textbf{0.212} 
&& \underline{\textbf{0.846}} & \textbf{2.615} & \underline{\textbf{0.577}} 
& \textbf{3.308} & \textbf{0.038}
& \textbf{3.577} & \underline{\textbf{0.346}} & \textbf{3.769} & \underline{\textbf{0.808}} & \underline{\textbf{4.269}} \\
        \bottomrule
      \end{tabular}
    \begin{tablenotes}
        \setlength\labelsep{0pt}
	\begin{scriptsize}
	\item
          The column ``n'' corresponds to the number of hold-out testing patients.
          The bset performance of each method is in \textbf{bold}.
          The best performance under each evaluation metric is \underline{underlined}.
	\end{scriptsize}
    \end{tablenotes}
  \end{threeparttable}
  \end{small}
\end{table*}

\begin{table*}[t]
  \centering
  \caption{Overall Performance in \LOV on 52 testing patients} 
  \label{table:lov:52}
  \begin{small}
  \begin{threeparttable}
    \begin{tabular}[]{
        @{\hspace{2pt}}l@{\hspace{2pt}}
        %
        %
        @{\hspace{3pt}}c@{\hspace{3pt}}
	@{\hspace{2pt}}c@{\hspace{1pt}}
        @{\hspace{1pt}}c@{\hspace{1pt}}
        @{\hspace{3pt}}c@{\hspace{3pt}}
        @{\hspace{2pt}}c@{\hspace{5pt}}
        @{\hspace{3pt}}c@{\hspace{3pt}}
        @{\hspace{2pt}}c@{\hspace{5pt}}
        @{\hspace{3pt}}c@{\hspace{3pt}}
        @{\hspace{2pt}}c@{\hspace{5pt}}
        @{\hspace{3pt}}c@{\hspace{3pt}}
        @{\hspace{2pt}}c@{\hspace{5pt}}
        @{\hspace{3pt}}c@{\hspace{3pt}}
	@{\hspace{2pt}}c@{\hspace{5pt}}
        @{\hspace{3pt}}c@{\hspace{2pt}}
        }
	\toprule
	\multirow{2}{*}{method} 
	& \multicolumn{2}{c}{feature level} && \multicolumn{10}{c}{task level} \\
	\cmidrule{2-3} \cmidrule{5-14}
	& QH@5 & WQH@5
        && NH$_1$@1 & NH$_1$@5 & NH$_2$@1 & NH$_2$@5
        & NH$_3$@1 & NH$_3$@5 & NH$_5$@1 & NH$_5$@5
        & NH$_{\text{all}}$@1 & NH$_{\text{all}}$@5 \\
        \midrule
	\multirow{6}{*}{\MTCRP}
& \textbf{1.385} & \textbf{1.668} 
&& 0.788 & 3.212 & 0.423 & 3.654 & 0.115 & 3.750 & 0.288 & 3.423 & 0.788 & 3.423 \\
& 1.327 & 1.616 
&& \underline{\textbf{0.808}} & \underline{\textbf{3.269}} & 0.423 & 3.654 & 0.115 & 3.731 & 0.173 & 3.423 & 0.788 & 3.404 \\
        %
&1.327 & 1.652 
&& 0.788 & 3.212 & \textbf{0.423} & \underline{\textbf{3.712}} & \textbf{0.115} & \underline{\textbf{3.750}} & 0.269 & 3.423 & 0.788 & 3.404 \\
%
& 1.308 & 1.616 
&& 0.788 & 3.154 & 0.423 & 3.654 & 0.115 & 3.712 & \textbf{0.288} & 3.481 & 0.788 & 3.615 \\
& 1.288 & 1.581 
&& 0.808 & 3.173 & 0.423 & 3.596 & 0.115 & 3.750 & 0.192 & \textbf{3.519} & 0.788 & 3.635 \\
& 1.269 & 1.616 
&& 0.808 & 3.115 & 0.423 & 3.635 & 0.115 & 3.731 & 0.250 & 3.481 & \underline{\textbf{0.788}} & \textbf{3.635} \\
	\midrule
	\multirow{7}{*}{\MTCRPH}
& \underline{\textbf{1.404}} & 1.656 
&& 0.750 & 2.827 & \textbf{0.404} & 3.250 & 0.173 & 3.481 & \underline{\textbf{0.385}} & 3.596 & \underline{\textbf{0.788}} & \underline{\textbf{4.154}} \\
& 1.365 & \underline{\textbf{1.695}}
&& 0.731 & 2.808 & 0.365 & 3.308 & 0.173 & 3.462 & 0.365 & 3.596 & 0.788 & 4.154 \\
& 1.327 & 1.562
&& \underline{\textbf{0.808}} & 3.077 & 0.404 & 3.365 & 0.135 & 3.577 & 0.250 & 3.673 & 0.788 & 4.115 \\
& 1.327 & 1.605 
&& 0.769 & \textbf{3.154} & 0.385 & \textbf{3.596} & 0.135 & 3.712 & 0.212 & 3.519 & 0.788 & 3.577 \\
& 1.308 & 1.609
&& 0.769 & 2.904 & 0.385 & 3.308 & \underline{\textbf{0.192}} & 3.442 & 0.365 & 3.654 & 0.788 & 4.154 \\
& 1.327 & 1.605
&& 0.769 & 3.154 & 0.385 & 3.596 & 0.135 & \textbf{3.712} & 0.212 & 3.519 & 0.788 & 3.577 \\
& 1.288 & 1.545
&& 0.788 & 3.000 & 0.404 & 3.385 & 0.154 & 3.558 & 0.308 & \textbf{3.712} & 0.788 & 4.154 \\
	\midrule
	\multirow{4}{*}{\KRL}
& \textbf{1.173} & \textbf{1.548}
&& 0.096 & 2.577 & 0.385 & \textbf{3.231} & 0.077 & \textbf{3.385} & \textbf{0.346} & \underline{\textbf{3.808}} 
& \underline{\textbf{0.788}} & \underline{\textbf{4.154}} \\
& 1.173 & 1.534
&&\textbf{0.154} & \textbf{2.615} & 0.250 & 3.192 & 0.077 & 3.385 & 0.346 & 3.712 & 0.788 & 4.154 \\
& 1.096 & 1.437 
&& 0.077 & 2.577 & \textbf{0.462} & 3.231 & 0.077 & 3.385 & 0.346 & 3.808 & 0.788 & 4.154 \\
& 0.423 & 0.504
&& 0.019 & 2.019 & 0.038 & 2.500 & \textbf{0.115} & 2.481 & 0.115 & 2.712 & 0.019 & 2.673 \\
	\midrule
	\BMTMKL
	& \textbf{0.403} & \textbf{0.255} 
	&& \underline{\textbf{0.808}} & \textbf{2.577} & \underline{\textbf{0.481}} 
	& \textbf{3.231} & \textbf{0.077} 
	& \textbf{3.385} & \textbf{0.346} & \textbf{3.596} & \underline{\textbf{0.788}} & \underline{\textbf{4.154}} \\
        \bottomrule
      \end{tabular}
    \begin{tablenotes}
        \setlength\labelsep{0pt}
	\begin{scriptsize}
	\item
          The column ``n'' corresponds to the number of hold-out testing patients. 
          The best performance of each model is in \textbf{bold}.
          The best performance under each evaluation metric is upon \underline{underline}.
	\end{scriptsize}
    \end{tablenotes}
  \end{threeparttable}
  \end{small}
\end{table*}

%
Table~\ref{table:lov:26} and Table~\ref{table:lov:52} present the performance of \MTCRP, \MTCRPH 
and two baseline methods in the \LOV setting.
{Due to space limit, we did not present the standard deviations in the tables, but they have similar trends
as those in Table~{\ref{table:cv}}. }
We first hold out 26 (Table~\ref{table:lov:26}) and 52 (Table~\ref{table:lov:52}) 
AD patients as testing patients, respectively. 
We determine these hold-out AD patients as the ones that have more than 10 similar
AD patients in the training set with corresponding patient similarities higher than 0.67 and 0.62, respectively.
%
%
%

%
\subsubsection{Comparison on cognitive feature level}
\label{sec:results:overall:lov:feature}
Table~\ref{table:lov:26} and Table~\ref{table:lov:52} show that \MTCRP and \MTCRPH significantly outperform the baseline methods 
in terms of all the evaluation metrics on cognitive feature level (i.e., QH@5 and WQH@5), 
which is consistent with the experimental results in \CV setting.
When 26 patients are hold out for testing,
with parameters $\alpha = 0.5$, $\beta = 1.5$, $\gamma = 1.0$ and $\text{d} = 30$, 
\MTCRP outperforms the best baseline method \KRL at 13.4\% and 1.3\% on QH@5 and WQH@5, respectively.
The performance of \MTCRPH is very comparable with that of \MTCRP" 
\MTCRPH outperforms \KRL at 13.4\% and 0.5\% on QH@5 and WQH@5, respectively.
When 52 patients are hold out for testing,
with parameters $\alpha = 0.5$, $\beta = 0.5$, $\gamma = 1.0$ and $\text{d} = 50$,
\MTCRP outperforms the best baseline method \KRL at 18.1\% and 7.8\% on QH@5 and WQH@5, respectively.
\MTCRPH even performs better than \MTCRP in this setting. In addition, 
\MTCRPH outperforms \KRL at 19.7\% and 9.5\% 
on QH@5 and WQH@5, respectively.
These experimental results demonstrate that for new patients, 
\MTCRP and \MTCRPH are able to rank more relevant features 
to the top of the ranking list than the two baseline methods. 
They also indicate that for new patients, ranking based methods (e.g., \MTCRP and \MTCRPH) are more effective 
than regression based methods (e.g., \KRL and \BMTMKL) for biomarker prioritization. 
%

\subsubsection{Comparison on cognitive task level}
\label{sec:results:overall:lov:task}
%
Table~\ref{table:lov:26} also shows that 
when 26 patients are hold out for testing, \MTCRP and \MTCRPH are both able to identify the top most relevant questionnaire for
84.6\% of the testing patients (i.e., 22 patients) under NH$_1@1$.
Table~\ref{table:lov:52} shows that
when 52 patients are hold out for testing, \MTCRP and \MTCRPH are both able to identify for 80.8\% of the testing patients
(i.e., 42 patients) under NH$_1@1$.
Note that the hold-out testing patients in \LOV do not have any cognitive features. Therefore,
the performance of \MTCRP and \MTCRPH as above demonstrates their strong capability in identifying most
AD related cognitive features based on imaging features only. 
We also find that \MTCRP and \MTCRPH are able to achieve similar or even better results compared to baseline methods
in terms of the evaluation metrics on cognitive task level (i.e., NH$_g$@1 and NH$_g$@5).
When 26 patients are hold out for testing,
\MTCRP and \MTCRPH outperform the baseline methods in terms of NH$_g$@1 (i.e., $g = 1, 2 \ldots 5$).
They are only slightly worse than \KRL on ranking relevant tasks 
on their top-5 of predictions when $g = 1$ or $g = 5$ (3.308 vs 3.423 on NH$_1$@5 and 3.808 vs 3.962 on NH$_5$@5).
When 52 patients are hold out for testing,
\MTCRP and \MTCRPH also achieve the best performance on most of the evaluation metrics.
They are only slightly worse than \KRL on NH$_2$@1, NH$_5$@5 (0.423 vs 0.481 on NH$_2$@1 and 3.712 vs 3.808 on NH$_5$@5).
%
These experimental results demonstrate that among top 5 tasks in the ranking list, \MTCRP and \MTCRPH 
rank more relevant task on top than \KRL. 
%

It's notable that in Table~\ref{table:lov:26} and Table~\ref{table:lov:52}, as the number of features used to score cognitive tasks (i.e.,
$g$ in NH$_g@k$) increases, the performance of all the methods in NH$_g@1$ first declines and then increases.
This may indicate that as $g$ increases, irrelevant features which happen to have relatively high scores will be included 
in scoring tasks, and thus degrade the model performance on NH$_g@1$. 
However, generally, the scores of irrelevant features are considerably lower than those of relevant ones. 
Thus, as more features are included, the scores for tasks are more dominated by the scores of relevant features and thus the performance increases.

We also find that \BMTMKL performs poorly on NH$_3@1$ in both Table~\ref{table:lov:26} and Table~\ref{table:lov:52}. 
This indicates that \BMTMKL, a regression-based method, could not well rank relevant features and irrelevant features. 
%
It's also notable that generally the best performance for the 26 testing patients is better than that for 52 testing patients.
This may be due to that the similarities between the 26 testing patients and their top 10 similar training patients are 
higher than those for the 52 testing patients.
The high similarities enable accurate latent vectors for testing patients. 

Table~\ref{table:lov:26} and Table~\ref{table:lov:52} also show that
\MTCRPH is better than \MTCRP on ranking cognitive tasks in \LOV setting.
When 26 patients are hold out for testing,
\MTCRPH outperforms \MTCRP on NH$_1$@5, NH$_5$@5 and NH$_\text{all}$@5 and achieves
very comparable performance on the rest metrics.
When 52 patients are hold out for testing,
\MTCRPH is able to achieve better performance than \MTCRP on QH@5, WQH@5,
NH$_3$@1, NH$_5$@1, NH$_5$@5 and NH$_\text{all}$@5 and also achieves very comparable performance on
the rest metrics.
%
Generally, \MTCRPH outperforms \MTCRP in terms of  metrics on cognitive task level. 
This demonstrates the effectiveness of hinge loss-based methods in separating relevant and irrelevant 
features during modeling. 
%

\section{Discussion}
\label{sec:results:case}

Our experimental results show that 
when NH$_1@1$ achieves its best performance of 0.846
for the 26 testing patients in the \LOV setting (i.e.,
the first row block in Table~\ref{table:lov:26}), 
the task that is most commonly prioritized for the testing patients is
Rey Auditory Verbal Learning Test (RAVLT), including the following cognitive features:
1) trial 1 total number of words recalled;
2) trial 2 total number of words recalled; 
3) trial 3 total number of words recalled; 
4) trial 4 total number of words recalled; 
5) trial 5 total number of words recalled; 
6) total Score; 
7) trial 6 total number of words recalled; 
8) list B total number of words recalled; 
9) 30 minute delay total; and 
10) 30 minute delay recognition score. 
RAVLT is also the most relevant task in the ground truth if tasks
are scored correspondingly. 
RAVLT assesses learning and memory, and has shown promising
performance in early detection of AD~\cite{Moradi2017}. A number of studies have reported high correlations
between various RAVLT scores with different brain regions~\cite{Balthazar2010}.
For instance, RAVLT recall is associated with medial prefrontal cortex and hippocampus;
RAVLT recognition is highly correlated with thalamic and caudate nuclei.
In addition, genetic analysis of $APOE$ $\varepsilon$4 allele, the most common variant of AD, 
reported its association with RAVLT score in an early-MCI (EMCI) study \cite{Risacher2013}.
The fact that RAVLT is prioritized demonstrates that \MTCRP is powerful in prioritizing
cognitive features to assist AD diagnosis. 

%
Similarly, we find the top-5 most frequent cognitive tasks corresponding to the performance at
NH$_3@5$=3.731 for the 26 hold-out testing patients. They are: Functional Assessment Questionnaire (FAQ), 
Clock Drawing Test (CDT), Weschler's Logical Memory Scale (LOGMEM), 
Rey Auditory Verbal Learning Test (RAVLT), and Neuropsychiatric Inventory Questionnaire (NPIQ). 
In addition to RAVLT discussed above, other top prioritized cognitive tasks 
have also been reported to be associated with AD or its progression. In an MCI to AD conversion study, FAQ, NPIQ and RAVLT 
showed significant difference between MCI-converter and MCI-stable groups~\cite{Risacher2009}.
We also notice that for some testing subjects, \MTCRP is able to  
very well reconstruct their ranking structures. 
For example, when NH$_3@5$ achieves its optimal performance 3.731, for a certain testing subject, 
her top-5 predicted cognitive tasks RAVLT, LOGMEM, FAQ, NPIQ and CDT are exactly the top-5 
cognitive tasks in the ground truth.
%
These evidences further demonstrate the diagnostic power of our method. 

\section{Conclusions}
\label{sec:disc}

We have proposed a novel machine learning paradigm to prioritize cognitive assessments based on their relevance to 
AD at the individual patient level. The paradigm tailors the cognitive biomarker discovery and cognitive assessment selection 
process to the brain morphometric characteristics of each individual patient. It has been implemented using newly developed
 learning-to-rank method \MTCRP and \MTCRPH.
Our empirical study on the ADNI data has produced promising results to identify and prioritize individual-specific cognitive biomarkers 
as well as cognitive assessment tasks based on the individual's structural MRI data.
In addition, \MTCRPH shows better performance than \MTCRP on ranking cognitive assessment tasks.
The resulting top ranked cognitive biomarkers and assessment tasks have the potential to aid personalized diagnosis and 
disease subtyping, and to make progress towards enabling precision medicine in AD. 

\section{Abbreviations}
\label{sec:abbr}
\textbf{AD}: Alzheimer's Disease

\noindent
\textbf{MRI}: Magnetic Resonance Imaging

\noindent
\textbf{ADNI}: Alzheimer's Disease Neuroimaging Initiative

\noindent
\textbf{LETOR}: Learning-to-Rank

\noindent
\textbf{PET}: Positron Emission Tomography

\noindent
\textbf{MCI}: Mild Cognitive Impairment

\noindent
\textbf{HC}: Healthy Control

\noindent
\textbf{ADAS}: Alzheimer's Disease Assessment Scale

\noindent
\textbf{CDR}: Clinical Dementia Rating Scale

\noindent
\textbf{FAQ}: Functional Assessment Questionnaire 

\noindent
\textbf{GDS}: Geriatric Depression Scale

\noindent
\textbf{MMSE}: Mini-Mental State Exam

\noindent
\textbf{MODHACH}: Modified Hachinski Scale

\noindent
\textbf{NPIQ}: Neuropsychiatric Inventory Questionnaire

\noindent
\textbf{BNT}: Boston Naming Test

\noindent
\textbf{CDT}: Clock Drawing Test

\noindent
\textbf{DSPAN}: Digit Span Test

\noindent
\textbf{DSYM}: Digit Symbol Test

\noindent
\textbf{FLUENCY}: Category Fluency Test

\noindent
\textbf{LOGMEM}: Weschler's Logical Memory Scale

\noindent
\textbf{RAVLT}: Rey Auditory Verbal Learning Test

\noindent
\textbf{TRAIL}: Trail Making Test

\noindent
\textbf{RBF}: Radial Basis Function

\noindent
\textbf{PLTR}: Joint Push and Learning-to-Rank Method

\noindent
\textbf{PLTR$_\textbf{h}$}: Joint Push and Learning-to-Rank Method using Hinge Loss

\noindent
\textbf{BMTMKL}: Bayesian Multi-Task Multi-Kernel Learning

\noindent
\textbf{KRL}: Kernelized Rank Learning

\noindent
\textbf{CV}: Cross Validation

\noindent
\textbf{LOV}: Leave-Out Validation

\noindent
\textbf{QH@$k$}: Average Feature Hit at $k$

\noindent
\textbf{WQH@$k$}: Weighted Average Feature Hit at $k$

\noindent
\textbf{$\text{NH}_g$@$k$}: Task Hit at $k$

\noindent
\textbf{APOE}: Apolipoprotein E

\noindent
\textbf{EMCI}: Early-MCI
\section{Declaration}
\label{sec:dec}

\subsection{Ethics approval and consent to participate}

The dataset supporting the conclusions of this article is available in the 
Alzheimer’s Disease Neuroimaging Initiative (ADNI)~\cite{adni}.
ADNI data can be requested by all interested investigators; they can request it via the ADNI website 
and must agree to acknowledge ADNI and its funders in the papers that use the data. 
There are also some other reporting requirements; the PI must give an annual report of what the data have been used for, 
and any publications arising.
More details are available at http://adni.loni.usc.edu/data-samples/access-data/. 

%

\subsection{Consent for publication}
Not applicable

\subsection{Availability of data and materials}
\label{sec:dec:data}
The dataset supporting the conclusions of this article is available in the 
Alzheimer’s Disease Neuroimaging Initiative (ADNI)~\cite{adni}.

\subsection{Conflicts of interest}
\label{sec:dec:cof}
The authors declare that they have no competing interests. 

\subsection{Funding}
\label{sec:dec:fund}
This work was supported in part by NIH R01 EB022574,  
R01 AG019771, and P30 AG010133; NSF 
IIS 1837964 and 1855501.
Any opinions, findings, and conclusions or recommendations
expressed in this material are those of the authors and do
not necessarily reflect the views of the funding agencies.

\subsection{Authors' contributions}
\label{sec:dec:contributions}

XN and LS designed the research study. BP and XY contributed to the
conduct of the study: XY extracted and processed the data from ADNI; BP conduced 
the model development and data analysis. 
The results were analyzed, interpreted and
discussed by BP, XY, SL, AJ, LS and XN. BP and XN drafted the manuscript and all
co-authors revised and approved the final version of the manuscript.

\subsection{Acknowledgements}
\label{sec:dec:ack}

Data collection and sharing for this project was funded by the Alzheimer's Disease Neuroimaging Initiative (ADNI) 
(National Institutes of Health Grant U01 AG024904) and 
DOD ADNI (Department of Defense award number W81XWH-12-2-0012). 
ADNI is funded by the National Institute on Aging, the National Institute of Biomedical Imaging and Bioengineering, 
and through generous contributions from the following: AbbVie, Alzheimer's Association; Alzheimer's Drug Discovery Foundation; 
Araclon Biotech; BioClinica, Inc.; Biogen; Bristol-Myers Squibb Company; CereSpir, Inc.; Eisai Inc.; Elan Pharmaceuticals, Inc.; 
Eli Lilly and Company; EuroImmun; F. Hoffmann-La Roche Ltd and its affiliated company Genentech, Inc.; Fujirebio; 
GE Healthcare; IXICO Ltd.; Janssen Alzheimer Immunotherapy Research \& Development, LLC.; Johnson \& Johnson Pharmaceutical Research
 \& Development LLC.; Lumosity; Lundbeck; Merck \& Co., Inc.; Meso Scale Diagnostics, LLC.; NeuroRx Research; Neurotrack Technologies; 
 Novartis Pharmaceuticals Corporation; Pfizer Inc.; Piramal Imaging; Servier; Takeda Pharmaceutical Company; and Transition Therapeutics. 
 The Canadian Institutes of Health Research is providing funds to support ADNI clinical sites in Canada. Private sector contributions 
 are facilitated by the Foundation for the National Institutes of Health (www.fnih.org). The grantee organization is the Northern California 
 Institute for Research and Education, and the study is coordinated by the Alzheimer's Disease Cooperative Study at the University of California, 
 San Diego. ADNI data are disseminated by the Laboratory for Neuro Imaging at the University of Southern California.

\bibliographystyle{bmc-mathphys}
\bibliography{paper}

\end{document}


\begin{frontmatter}

\begin{fmbox}
\dochead{Research}


\title{Cognitive Biomarker Prioritization in Alzheimer's 
Disease using Brain Morphometric Data (Supplementary Materials)}

\author[
   addressref={aff1},                   
   email={peng.707@buckeyemail.osu.edu} 
]{\inits{BP}\fnm{Bo} \snm{Peng}}
\author[
   addressref={aff2},
   email={Xiaohui.Yao@pennmedicine.upenn.edu}
]{\inits{XY}\fnm{Xiaohui} \snm{Yao}}
\author[
   addressref={aff3},
   email={srisache@iupui.edu}
]{\inits{SL}\fnm{Shannon L.} \snm{Risacher}}
\author[
   addressref={aff3},
   email={asaykin@iupui.edu}
]{\inits{AJ}\fnm{Andrew J.} \snm{Saykin}}
\author[
   addressref={aff2},
   email={li.shen@pennmedicine.upenn.edu}
]{\inits{LS}\fnm{Li} \snm{Shen}}
\author[
   addressref={aff1},
   corref={aff1},
   email={ning.104@osu.edu}
]{\inits{XN}\fnm{Xia} \snm{Ning}}
\author[
  noteref={n2}
]{\inits{ADNI}\fnm{for the} \snm{ADNI}}


\address[id=aff1]{
  \orgname{The Ohio State University}, 
  \city{Columbus},                                  
  \cny{US}                                              
}
\address[id=aff2]{%
  \orgname{University of Pennsylvania},
  \city{Philadelphia},
  \cny{US}
}
\address[id=aff3]{%
  \orgname{Indiana University},
  \city{Indianapolis},
  \cny{US}
}


\begin{artnotes}
%
\note[id=n2]{Data used in preparation of this article were obtained from the Alzheimer’s Disease Neuroimaging 
Initiative (ADNI) database (adni.loni.usc.edu). As such, the investigators within the ADNI contributed to the design and 
implementation of ADNI and/or provided data but did not participate in analysis or writing of this report. A complete listing of 
ADNI investigators can be found at: https://adni.loni.usc.edu/wp-content/uploads/how\_to\_apply/ADNI\_Data\_Use\_Agreement.pdf. }
\end{artnotes}



%
\end{fmbox}

\section{Features used in experiments}
\label{sec:features}

Table~\ref{table:cog_features} presents the cognitive features that are used in our experiments after data processing. 
Table~\ref{table:img_features} presents the image features that are used in our experiments after data processing. 

\setcounter{table}{0}
\renewcommand{\thetable}{S\arabic{table}}

\onecolumn
\begin{center}
\begin{ThreePartTable}
\begin{TableNotes}
\item The "Task" column corresponds to the task that the cognitive feature belongs to. 
The "Measurement/Question" column corresponds to the measurement/question that the cognitive is.\par
\end{TableNotes}
\begin{longtable}[]{
      @{\hspace{1pt}}l@{\hspace{10pt}}
      @{\hspace{1pt}}p{0.75\linewidth}@{\hspace{1pt}}
              }
      \caption{Cognitive Features Used in Experiments} \label{table:cog_features}\\
      \hline
      Task & Measurement/Question\\ 
      \hline
      \endfirsthead

      \multicolumn{2}{c}%
      {{\bfseries \tablename\ \thetable{} -- continued from previous page}} \\
      \hline
      Task & Measurement/Question\\
      \hline
      \endhead
      \hline \multicolumn{2}{|r|}{{Continued on next page}} \\
      \endfoot
      \hline
      \insertTableNotes 
      \endlastfoot
      ADAS & Q1 - Word Recall\\
      ADAS & Q2 - Commands\\
      ADAS & Q3 - Construction\\
      ADAS & Q4 - Delayed Word Recall\\
      ADAS & Q5 - Naming\\
      ADAS & Q6 - Ideational Praxis\\
      ADAS & Q7 - Orientation\\
      ADAS & Q8 - Word Recognition\\
      ADAS & Q9 - Recall Instructions\\
      ADAS & Q10 - Spoken Language\\
      ADAS & Q11 - Word Finding\\
      ADAS & Q12 - Comprehension\\
      ADAS & Q14 - Number Cancellation\\
      ADAS & Sub-Total (classic 70 point total that excludes Q4 (Delayed Word Recall) and Q14 (Number Cancellation))\\
      ADAS & Total (85 point total that includes Q4 (Delayed Word Recall) and Q14 (Number Cancellation))\\
      CDR & CDR Memory score\\
      CDR & CDR Orientation score\\
      CDR & CDR Judgement and Problem Solving score\\
      CDR & CDR Community Affairs score\\
      CDR & CDR Home \& Hobbies score\\
      CDR & CDR Personal Care score\\
      CDR & Global CDR score\\
      CDR & CDR Sum-of-Boxes score\\
      FAQ & Any difficulty in writing checks paying bills and/or balancing checkbook?\\
      FAQ & Any difficulty in assembling tax records business affairs or other papers?\\
      FAQ & Any difficulty in shopping alone for clothes household necessities or groceries?\\
      FAQ & Any difficulty in playing a game of skill such as bridge or chess or working on a hobby?\\
      FAQ & Any difficulty in heating water making a cup of coffee or turing off the stove?\\
      FAQ & Any difficulty in preparing a balanced meal?\\
      FAQ & Any difficulty in keeping track of current events?\\
      FAQ & Any difficulty in paying attention to and understanding a TV program book or magazine?\\
      FAQ & Any difficulty in remembering appointments family occasions holidays medications?\\
      FAQ & Any difficulty in traveling out of the neighborhood driving or arranging to take public transportation?\\
      FAQ & Total FAQ score\\
      GDS & Q2 - Have you dropped many of your activities and interests?  (YES = 1 point)\\
      GDS & Q4 - Do you often get bored?  (YES = 1 point)\\
      GDS & Q8 - Do you often feel helpless?  (YES = 1 point)\\
      GDS & Q10 - Do you feel you have more problems with memory than most?  (YES = 1 point)\\
      GDS & Q15 - Do you think that most people are better off than you are?  (YES = 1 point)\\
      GDS & GDS Total Score\\
      MMSE & Q1 - What is today's date?\\
      MMSE & Q2 - What is the year?\\
      MMSE & Q3 - What is the month?\\
      MMSE & Q4 - What day of the week is today?\\
      MMSE & Q5 - What season is it?\\
      MMSE & Q6 - What is the name of this hospital (clinic place)?\\
      MMSE & Q7 - What floor are we on?\\
      MMSE & Q8 - What town or city are we in?\\
      MMSE & Q9 - What county (district borough area) are we in?\\
      MMSE & Q15 - L\\
      MMSE & Q16 - R\\
      MMSE & Q17 - O\\
      MMSE & Q18 - W\\
      MMSE & Q19 - Ball\\
      MMSE & Q20 - Flag\\
      MMSE & Q21 - Tree\\
      MMSE & Q24 - Say: Repeat after me: no ifs ands or buts.\\
      MMSE & Q30 - Present the participant with the Construction Stimulus page. Say: Copy this design.\\
      MMSE & MMSE Total Score\\
      NIP & QA - Does participant believe that others are stealing from him/her or planning to harm him/her in some way? (Delusions)\\
      NIP & QC - Is the participant stubborn and resistive to help from others? (Agitation/Aggression)\\
NIP
          & QD - Does the participant act as if he/she is sad or in low spirits? 
           Does he/she cry? (Depression/Dysphoria)\\
NIP
          & QE - Does the participant become upset when separated from you? Does he/she have any other signs of nervousness 
           such as shortness of breath sighing being unable to relax or feeling excessively tense? (Anxiety)\\
NIP          & QG - Does the participant seem less interested in his/her usual activities 
           and in the activities and plans of others? (Apathy/Indifference)\\
NIP
          & QH - Does the participant seem to act impulsively? For example does the participant talk to strangers 
           as if he/she knows them  or does the participant say things that may hurt people's feelings? (Disinhibition)\\
NIP          & QI - Is the participant impatient or cranky? Does he/she have difficulty coping 
          with delays or waiting for planned activities? (Irritability/Lability)\\
NIP          & QJ - Does the participant engage in repetitive activities such as pacing around the house handling buttons 
            wrapping strings  or doing other things repeatedly? (Abberant Motor Behavior)\\
NIP          & QK - Does the participant awaken you during the night rise too early in the morning 
          or take excessive naps during the day? (Sleep)\\
      NIP & QL - Has the participant lost or gained weight or had a change in the food he/she likes? (Appetite)\\
      NIP & Total NPIQ score\\
      BNT & Baseline total ANART score (Total number of errors on the ANART)\\
      BNT & Boston Naming Test - Number of spontaneously given correct responses\\
      BNT & Boston Naming Test - Number of semantic cues given\\
      BNT & Boston Naming Test - Number of phonemic cues given\\
      BNT & Boston Naming Test - Number of correct responses following a phonemic cue\\
      BNT & Boston Naming Test Total Score \\
      CLOCK & Clock Drawing - Symmetry of number placement\\
      CLOCK & Clock Drawing - Correctness of numbers\\
      CLOCK & Clock Drawing - Presence of the two hands\\
      CLOCK & Clock Drawing - Presence of the two hands set to ten after eleven\\
      CLOCK & Clock Drawing Total Score\\
      CLOCK & Clock Copy - Symmetry of number placement\\
      CLOCK & Clock Copy - Correctness of numbers\\
      CLOCK & Clock Copy - Presence of the two hands set to ten after eleven\\
      CLOCK & Clock Copy Total Score\\
      DIGITSPAN & Digit Span Forward - Total Correct\\
      DIGITSPAN & Digit Span Forward - Length\\
      DIGITSPAN & Digit Span Backward - Total Correct\\
      DIGITSPAN & Digit Span Backward - Length\\
      DIGITSYM & Digit Symbol Total Score\\
      FLUENCY & Category Fluency (Animals) - Total Correct\\
      FLUENCY & Category Fluency (Animals) - Perseverations\\
      FLUENCY & Category Fluency (Vegetables) - Total Correct\\
      FLUENCY & Category Fluency (Vegetables) - Perseverations\\
      FLUENCY & Category Fluency (Vegetables) - Intrusions\\
      LOGMEM & Wechler's Logical Memory Immediate Recall - Total Number of Story Units Recalled\\
      LOGMEM & Wechler's Logical Memory Delayed Recall - Total Number of Story Units Recalled\\
      RAVLT & Trial 1 Total Number of Words Recalled\\
      RAVLT & Trial 2 Total Number of Words Recalled\\
      RAVLT & Trial 3 Total Number of Words Recalled\\
      RAVLT & Trial 4 Total Number of Words Recalled\\
      RAVLT & Trial 5 Total Number of Words Recalled\\
      RAVLT & RAVLT Total Score\\
      RAVLT & Trial 6 Total Number of Words Recalled\\
      RAVLT & List B Total Number of Words Recalled\\
      RAVLT & 30 Minute Delay Total\\
      RAVLT & 30 Minute Delay Recognition Score\\
      TRAIL & Trails A - Time to Complete (seconds)\\
      TRAIL & Trails A - Errors of Omission\\
      TRAIL & Trails B - Time to complete (seconds)\\
      TRAIL & Trails B - Errors of Omission\\
      TRAIL & Trails B - Errors of Commission\\
\end{longtable}
\end{ThreePartTable}
\end{center}


\onecolumn
\begin{center}
\begin{ThreePartTable}
\begin{TableNotes}
\item The "Feature" column corresponds to the feature name. 
The "Description" column corresponds to the description of the feature.\par
\end{TableNotes}
\begin{longtable}[]{
      @{\hspace{1pt}}l@{\hspace{10pt}}
      @{\hspace{1pt}}p{0.7\textwidth}@{\hspace{1pt}}
              }
      \caption{Baseline Imaging Features Used in Experiments.} \label{table:img_features}\\
      \hline
      Feature & Description\\
      \hline
      \endfirsthead

      \multicolumn{2}{c}%
      {{\bfseries \tablename\ \thetable{} -- continued from previous page}} \\
      \hline
      Feature & Description\\
      \hline
      \endhead
      \hline \multicolumn{2}{|r|}{{Continued on next page}} \\
      \endfoot
      \hline
      \insertTableNotes
      \endlastfoot
LCerebWM	&	Left Cerebral White Matter Volume	\\
LCerebCtx	&	Left Cerebral Cortex Volume	\\
RCerebCtx	&	Right Cerebral Cortex Volume	\\
LLatVent	&	Left Lateral Ventricle Volume	\\
RLatVent	&	Right Lateral Ventricle Volume	\\
LInfLatVent	&	Left Inferior Lateral Ventricle Volume	\\
RInfLatVent	&	Right Inferior Lateral Ventricle Volume	\\
LPutamVol	&	Left Putamen Volume	\\
RPutamVol	&	Right Putamen Volume	\\
LHippVol	&	Left Hippocampus Volume	\\
RHippVol	&	Right Hippocampus Volume	\\
LAmygVol	&	Left Amygdala Volume	\\
RAmygVol	&	Right Amygdala Volume	\\
LAccumVol	&	Left Accumbens Volume	\\
RAccumVol	&	Right Accumbens Volume	\\
CSF	&	CSF Volume	\\
CC\_Post	&	Posterior Corpus Collosum Volume	\\
CC\_MidPost	&	Mid-Posterior Corpus Collosum Volume	\\
CC\_Cent	&	Central Corpus Collosum Volume	\\
CC\_MidAnt	&	Mid-Anterior Corpus Collosum Volume	\\
CC\_Ant	&	Anterior Corpus Collosum Volume	\\
LBanksSTS	&	Left Banks Of The Superior Temporal Sulcus Thickness	\\
RBanksSTS	&	Right Banks Of The Superior Temporal Sulcus Thickness	\\
LCaudAntCing	&	Left Caudal Anterior Cingulate Thickness	\\
RCaudAntCing	&	Right Caudal Anterior Cingulate Thickness	\\
LCaudMidFrontal	&	Left Caudal Middle Frontal Gyri Thickness	\\
RCaudMidFrontal	&	Right Caudal Middle Frontal Gyri Thickness	\\
RCuneus	&	Right Cuneus Thickness	\\
LEntCtx	&	Left Entorhinal Cortex Thickness	\\
REntCtx	&	Right Entorhinal Cortex Thickness	\\
LFusiform	&	Left Fusiform Gyri Thickness	\\
RFusiform	&	Right Fusiform Gyri Thickness	\\
LInfParietal	&	Left Inferior Parietal Gyri Thickness	\\
RInfParietal	&	Right Inferior Parietal Gyri Thickness	\\
LInfTemporal	&	Left Inferior Temporal Gyri Thickness	\\
RInfTemporal	&	Right Inferior Temporal Gyri Thickness	\\
LIsthmCing	&	Left Isthmus Cingulate Thickness	\\
RIsthmCing	&	Right Isthmus Cingulate Thickness	\\
LLatOccipital	&	Left Lateral Occipital Gyri Thickness	\\
RLatOccipital	&	Right Lateral Occipital Gyri Thickness	\\
LLatOrbFrontal	&	Left Lateral Orbitofrontal Gyri Thickness	\\
RLatOrbFrontal	&	Right Lateral Orbitofrontal Gyri Thickness	\\
LLingual	&	Left Lingual Gyri Thickness	\\
RLingual	&	Right Lingual Gyri Thickness	\\
LMedOrbFrontal	&	Left Medial Orbitofrontal Thickness	\\
RMedOrbFrontal	&	Right Medial Orbitofrontal Thickness	\\
LMidTemporal	&	Left Middle Temporal Gyri Thickness	\\
RMidTemporal	&	Right Middle Temporal Gyri Thickness	\\
LParahipp	&	Left Paracentral Lobule Thickness	\\
RParahipp	&	Right Paracentral Lobule Thickness	\\
LParacentral	&	Left Parahippocampal Gyri Thickness	\\
RParacentral	&	Right Parahippocampal Gyri Thickness	\\
LParsOper	&	Left Pars Opercularis Thickness	\\
RParsOper	&	Right Pars Opercularis Thickness	\\
LParsOrb	&	Left Pars Orbitalis Thickness	\\
RParsOrb	&	Right Pars Orbitalis Thickness	\\
LParsTriang	&	Left Pars Triangularis Thickness	\\
RParsTriang	&	Right Pars Triangularis Thickness	\\
LPostCent	&	Left Postcentral Gyri Thickness	\\
RPostCent	&	Right Postcentral Gyri Thickness	\\
LPostCing	&	Left Posterior Cingulate Thickness	\\
RPostCing	&	Right Posterior Cingulate Thickness	\\
LPrecent	&	Left Precentral Gyri Thickness	\\
RPrecent	&	Right Precentral Gyri Thickness	\\
LPrecuneus	&	Left Precuneus Thickness	\\
RPrecuneus	&	Right Precuneus Thickness	\\
LRostAntCing	&	Left Rostral Anterior Cingulate Thickness	\\
RRostAntCing	&	Right Rostral Anterior Cingulate Thickness	\\
LRostMidFrontal	&	Left Rostral Middle Frontal Gyri Thickness	\\
RRostMidFrontal	&	Right Rostral Middle Frontal Gyri Thickness	\\
LSupFrontal	&	Left Superior Frontal Gyri Thickness	\\
RSupFrontal	&	Right Superior Frontal Gyri Thickness	\\
LSupParietal	&	Left Superior Parietal Gyri Thickness	\\
RSupParietal	&	Right Superior Parietal Gyri Thickness	\\
LSupTemporal	&	Left Superior Temporal Gyri Thickness	\\
RSupTemporal	&	Right Superior Temporal Gyri Thickness	\\
LSupramarg	&	Left Supramarginal Gyri Thickness	\\
RSupramarg	&	Right Supramarginal Gyri Thickness	\\
LFrontalPole	&	Left Frontal Pole Thickness	\\
RFrontalPole	&	Right Frontal Pole Thickness	\\
LTemporalPole	&	Left Temporal Pole Thickness	\\
RTemporalPole	&	Right Temporal Pole Thickness	\\
LTransvTemporal	&	Left Transverse Temporal Pole Thickness	\\
RTransvTemporal	&	Right Transverse Temporal Pole Thickness	\\
LHippGM	&	Left Hippocampus Gray Matter Volume	\\
RHippGM	&	Right Hippocampus Gray Matter Volume	\\

\end{longtable}
\end{ThreePartTable}
\end{center}

\end{frontmatter}

\bibliographystyle{bmc-mathphys}
\bibliography{paper}
